\documentclass[]{spie}  

\usepackage[english]{babel} 
\usepackage{amsmath,amsfonts,stmaryrd,amssymb} 
\usepackage{graphicx}
\usepackage{url}
\usepackage[colorlinks=true, allcolors=blue]{hyperref}
\graphicspath{ {images/} }
\usepackage{subcaption}
\usepackage{authblk}

\begin{document}
\title{Proton Radiation Damage Experiment on a Hybrid CMOS Detector}
\author[1]{Evan Bray}
\author[1]{Abraham D. Falcone}
\author[1]{Mitchell Wages}
\author[1]{David N. Burrows}
\author[2]{Carl R. Brune}
\author[2]{Donald Carter}
\author[2]{Devon Jacobs}

\affil[1]{Pennsylvania State University}
\affil[2]{Ohio University}
\maketitle 

\begin{abstract}
We report on the initial results of an experiment to determine the effects of proton radiation damage on an X-ray hybrid CMOS detector (HCD). The device was irradiated at the Edwards Accelerator Lab at Ohio University with 8 MeV protons, up to a total absorbed dose of 3 krad(Si) (4.5 $\times$ 10$^9$ protons/cm$^2$). The effects of this radiation on read noise, dark current, gain, and energy resolution are then analyzed. This exposure is the first of several which will be used for characterizing detector performance at absorbed dose levels that are relevant for imaging devices operating in a deep-space environment.
\end{abstract}
\section{Introduction}
Since the presence of high energy protons can negatively impact the performance of space-based X-ray imagers, understanding their effects is vital in planning for the design and operation of future X-ray space telescopes. The effects of various types of radiation damage on solid-state imaging devices has been well documented in a number of methodical studies and simulations over the past several decades\cite{VirmontoisDec2011, Hopkinson1992, BasslerMar2010, HopkinsonDec2000, HopkinsonApril1996}, and an excellent review of the relevant physics is given in Srour 2003\cite{Srour2003}. The primary cause of performance degradation in modern solid-state image sensors operating in a space environment is displacement damage in the silicon lattice caused by high energy protons and electrons\cite{Janesick1989}. Although the general effects of radiation damage follow a characteristic trend, the precise details are largely a function of device architecture.

The detector used in this experiment is a HyViSi H1RG hybrid CMOS detector fabricated by Teledyne Imaging Sensors, which we are characterizing for the purposes of X-ray astronomy. It is composed of 1024$\times$1024 pixels with an 18$\mu$m pitch, and a 100$\mu$m thick silicon depletion layer. Recent progress in characterizing HCDs for X-ray astronomy applications are discussed in several recent papers\cite{Hull2017,Hull2018,Bray2018}.

\section{Experimental Setup}
\begin{figure}
  \centering
  \includegraphics[height = 6cm]{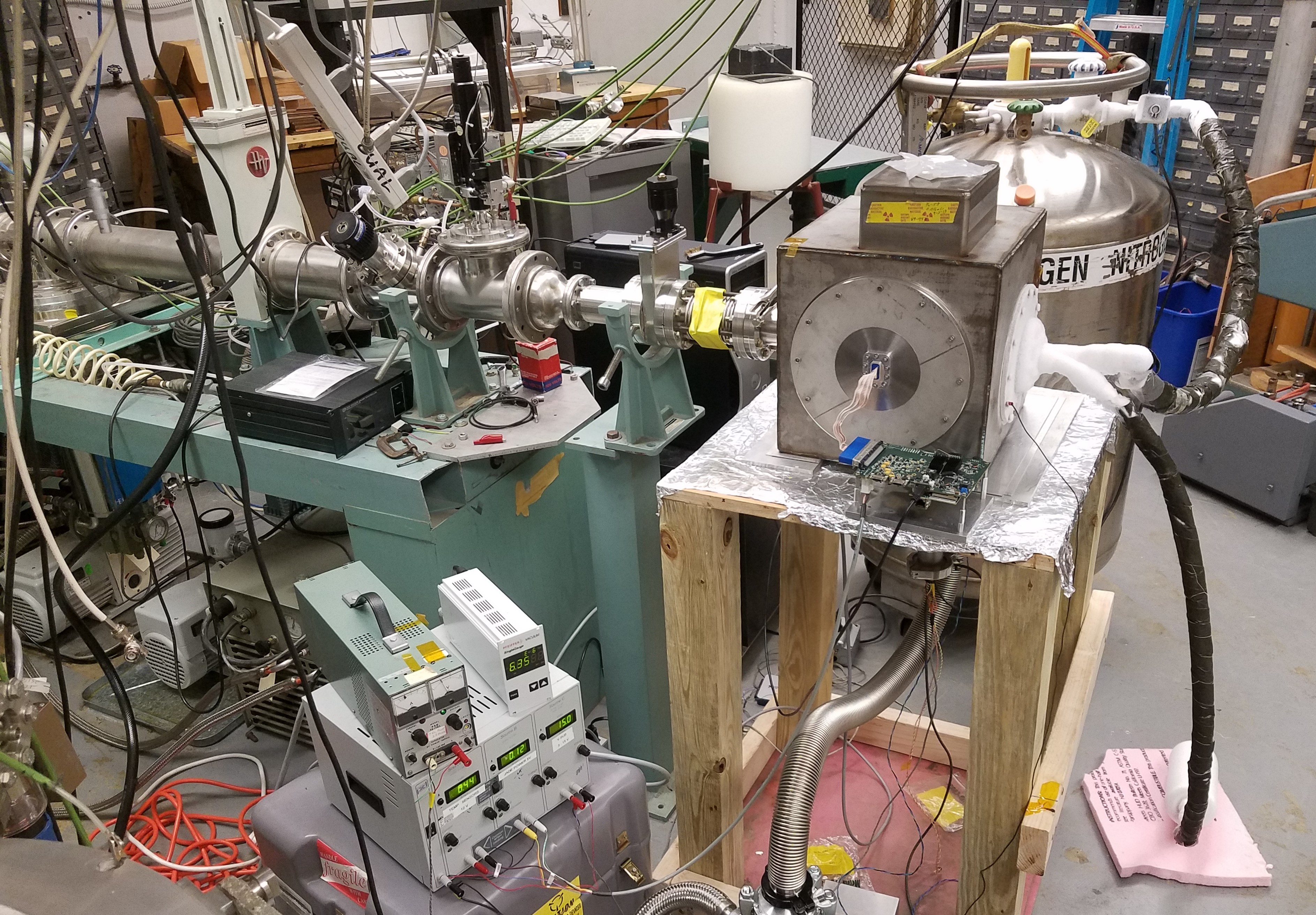}
  \caption{The ``Cube" vacuum chamber contains the Hybrid CMOS Detector, along with all necessary electronics and cooling hardware. It was precisely aligned and mounted to the tandem Van de Graff accelerator at Ohio University in preparation for irradiation.}
\label{fig:Radiation Experiment Setup}
\end{figure}

This experiment was conducted as part of a collaboration with the Edwards Accelerator Laboratory, operated by the Department of Physics and Astronomy at Ohio University. By mounting our detector setup on their tandem electrostatic accelerator, we are able to irradiate a controlled portion of the detector by utilizing a custom mask composed of thin tungsten sheets. The dose of 8-MeV protons was delivered over a period of $\sim$10 seconds and was measured using an upstream Faraday cup and beam-profile monitor that had been previously calibrated against the beam current and beam areal distribution measured at the detector position using a quartz window viewer. The detector was cooled to its operating temperature of 150K and left unbiased during this time. Exposure to X-rays was controlled by using a radioactive Iron-55 source that is mounted on an internal shutter system of the vacuum chamber. This allows us to switch between taking X-ray data and dark frames without breaking the primary vacuum. Figure \ref{fig:Radiation Experiment Setup} shows our ``Cube" vacuum chamber mounted on the end of the accelerator just before irradiation. Because the proton beam measures roughly 1cm $\times$ 2cm in this case, proper alignment and orientation of the Cube was of high priority, and was accomplished with the use of a theodolite.

\section{Post-irradiation Characterization}
After the HCD received a total absorbed dose of 3 krad(Si), we began the process of taking data at logarithmically spaced time intervals in order to characterize common detector properties over time. The properties to be examined in this fashion include read noise (Section \ref{sec:ReadNoise}), dark current (Section \ref{sec:DarkCurrent}), gain, and energy resolution (Section \ref{sec:EnergyResolution}). We also discuss some likely activation events in post-irradiation data that must be taken into account during data analysis (Section \ref{sec:Activation Events}).

After some amount of time at non-cryogenic temperatures, detector features are expected to change as the damaged silicon lattice ``repairs" itself through the process of annealing. After vacancies and defects are produced in the silicon, they naturally reorder themselves into more stable, less detrimental configurations. This process is often aided by baking out the detector at higher temperatures, and one study demonstrates that a 24 hour bake at 100$^{\circ}$C is equivalent to several months of storage at room temperature\cite{Hopkinson1992}. The effects and viability of annealing are still the subject of ongoing research, and have been shown to vary significantly between device architectures. For this reason, the only reliable way to determine the response of an imaging device to radiation damage is through direct experimentation. Our results indicate that the annealing process has caused detector characteristics to return very closely to its pre-irradiation levels.

\begin{table}[]
\centering
\begin{tabular}{lcccc}
\hline
 & \multicolumn{4}{c}{\begin{tabular}[c]{@{}c@{}}Median Detector Characteristics\\ (damaged region only)\end{tabular}} \\ \cline{2-5} 
Time After Irradiation & \begin{tabular}[c]{@{}c@{}}Read Noise \\ (e$^{-}$)\end{tabular} & \begin{tabular}[c]{@{}c@{}}Dark Current \\ (e$^{-}$/s/pixel)\end{tabular} & \begin{tabular}[c]{@{}c@{}}Gain \\ (DN/e$^{-}$)\end{tabular} & \begin{tabular}[c]{@{}c@{}}Energy Resolution \\ @ 5.9 keV\end{tabular} \\ \hline
\multicolumn{1}{l|}{Pre-irradiation} & 7.6 & .04 & 1.11 & 6.9$\%$ \\
\multicolumn{1}{l|}{8 hours} & 12.3 & .09 & 0.98 & 12.9$\%$ \\
\multicolumn{1}{l|}{32 hours} & 11.5 & .09 & 1.03 & 12.2$\%$ \\
\multicolumn{1}{l|}{104 hours} & 10.9 & .06 & 1.07 & 12.0$\%$ \\
\multicolumn{1}{l|}{3200 hours} & 7.7 & .04 & 1.12 & 6.9$\%$ \\ \hline
\end{tabular}
\caption{A quantitative summary of detector characteristics over time}
\label{table:summary}
\end{table}

\subsection{Read Noise} \label{sec:ReadNoise}
\begin{figure}
\begin{subfigure}{.50\textwidth}
  \centering
  \includegraphics[height = 6.0cm]{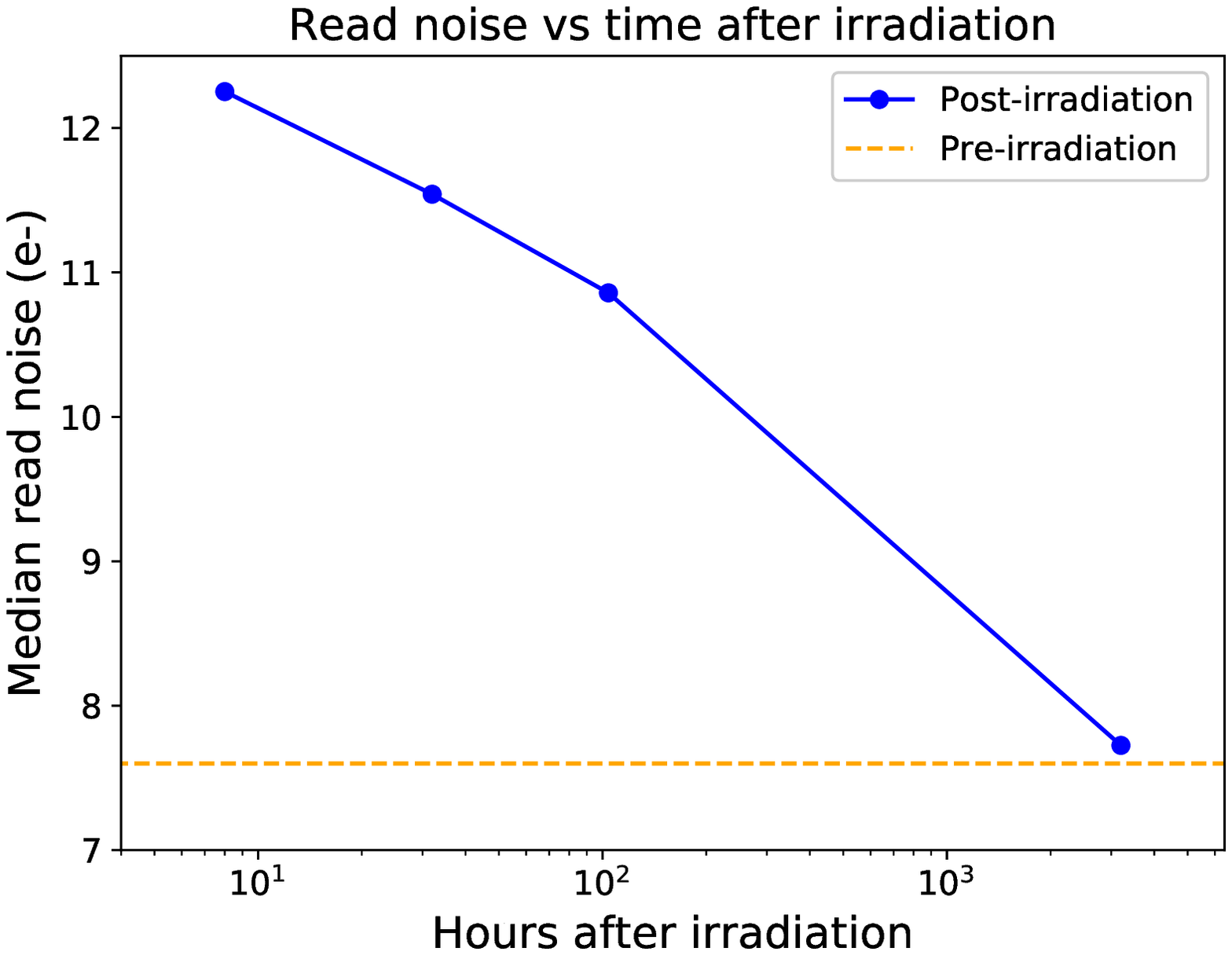}
\end{subfigure}
\begin{subfigure}{.49\textwidth}
	\centering
	\includegraphics[height = 6cm]{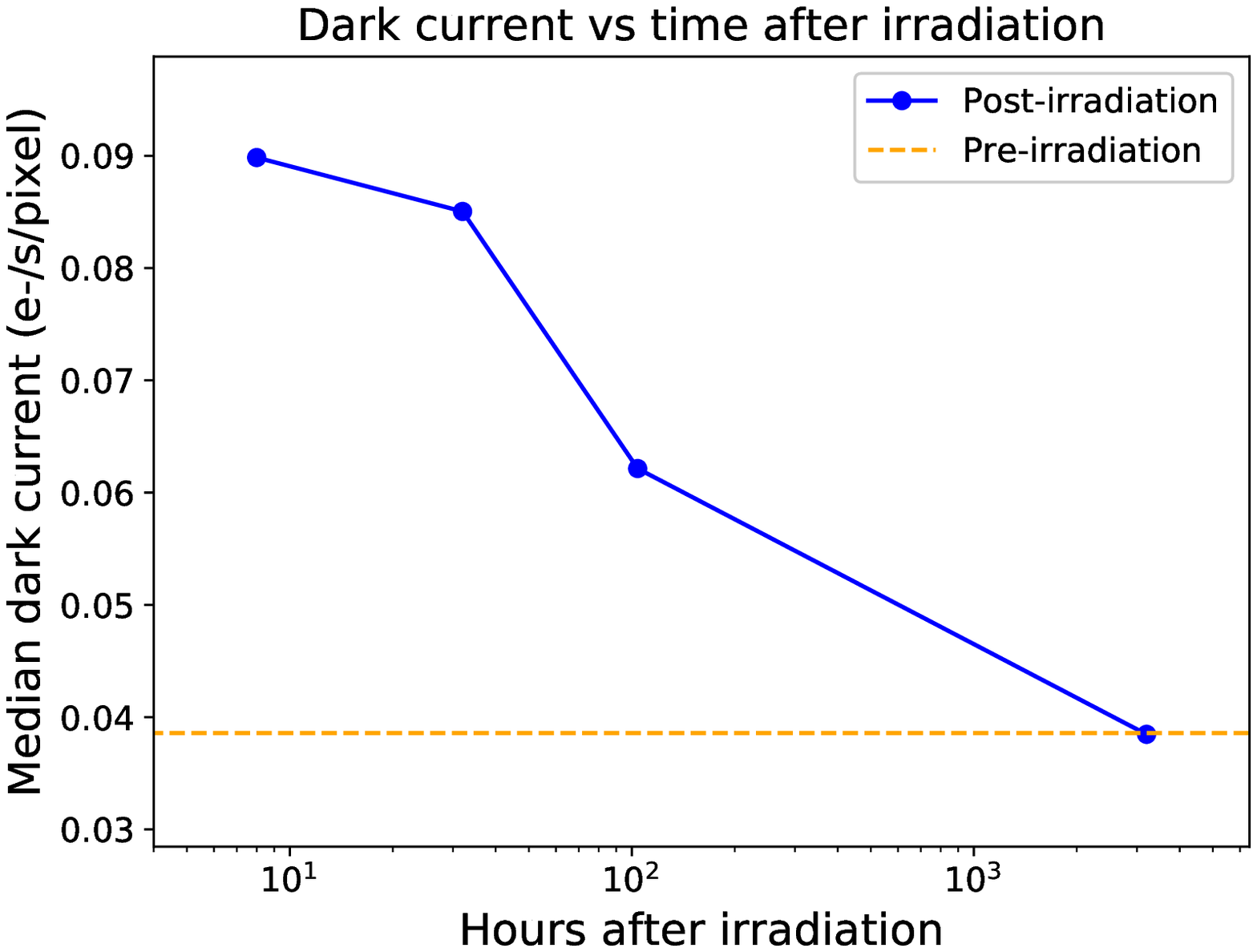}
\end{subfigure}
  \caption{(Left) A plot of median read noise vs time for the damaged region of the detector. (Right) A plot of median dark current vs time for the damaged region of the detector. After several months of storage at room temperature, both characteristics appear to have returned to pre-irradiation levels.}
\label{fig:characteristics_vs_time_plot}
\end{figure}

\begin{figure}
  \centering
  \includegraphics[height = 5.8cm]{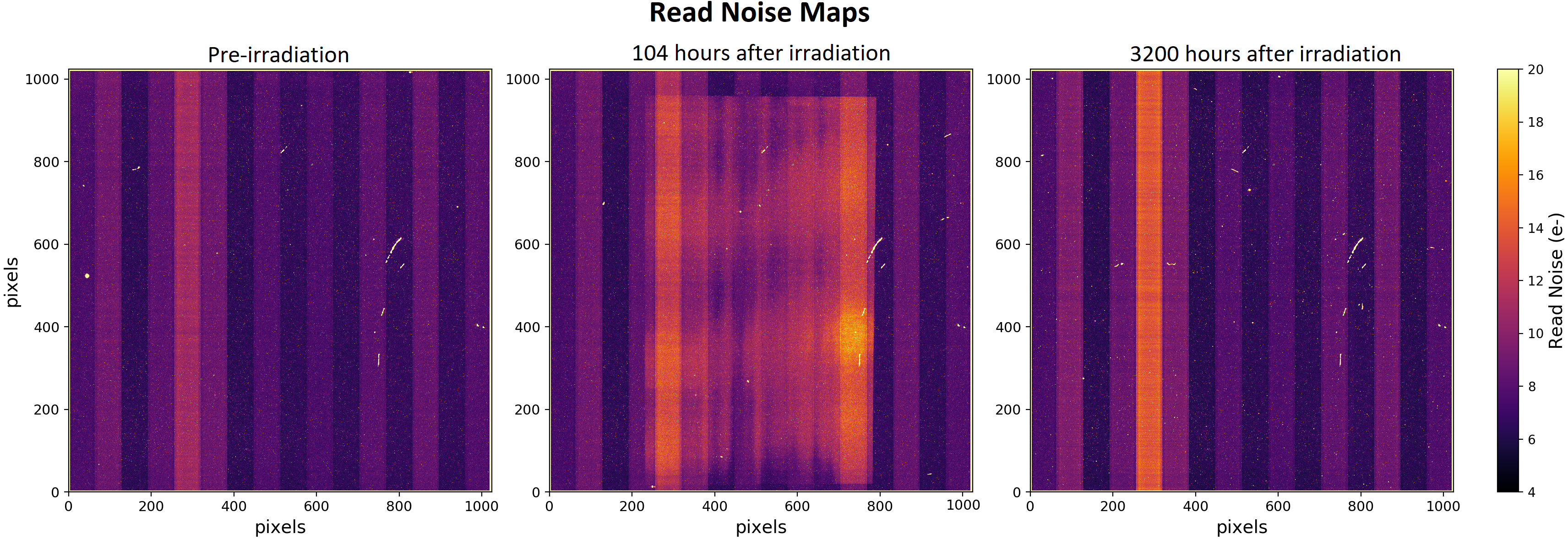}
  \caption{A map of the read noise across the detector as measured pre-irradiation (Left), 104 hours after irradiation (Middle), and 3200 hours afterward (Right). The damaged region can be seen in the middle, with the control regions on either side. Differences in read noise between channels appear as vertical striping in all images.}
\label{fig:read_noise_maps}
\end{figure}

\begin{figure}
  \centering
  \includegraphics[height = 9cm]{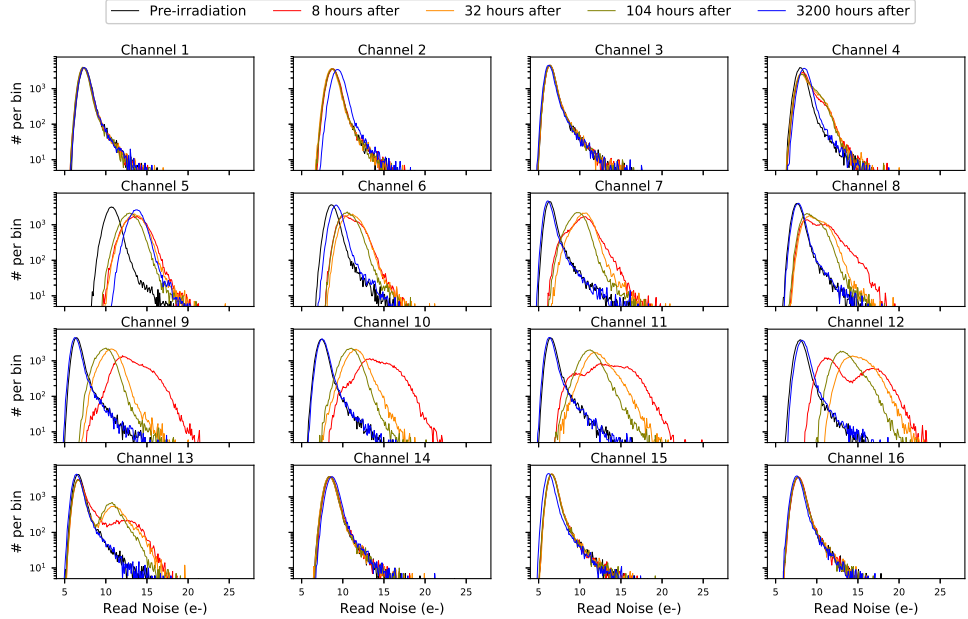}
  \caption{A read noise histogram of each readout channel is compared at multiple epochs after irradiation. When viewed in this way, the differences in read noise between channels become clear, and it can be seen that the read noise generally returns to pre-irradiation values by the time of the measurements at 3200 hours. Channels 5-12 were fully irradiated, while channels 1-4 and 13-16 were at least partially shielded by a tungsten mask. One exception to the norm is Channel 5, which is exhibiting uncharacteristically high read noise relative to the other channels pre-irradiation, as well as post-irradiation. This is likely to be related to the test stand electrical interface, as opposed to radiation damage.}
\label{fig:noise_vs_time_multichannel_hist}
\end{figure}

The read noise of the detector has been evaluated for each pixel within the damaged region over a large number of images at several epochs after irradiation. A plot of median read noise vs. time can be seen in Figure \ref{fig:characteristics_vs_time_plot}. By analyzing each pixel individually, we can identify regions of the detector that are inherently noisier than others, as shown in Figure \ref{fig:read_noise_maps}. Due to significant pre-existing differences in read noise between channels, along with potential nonuniformities in the proton beam, we find it insightful to also view a histogram of the read noise on a channel-by-channel basis, as shown in Figure \ref{fig:noise_vs_time_multichannel_hist}. After a large initial shift, read noise appears to have returned to pre-irradiation levels.

\subsection{Dark Current} \label{sec:DarkCurrent}
\begin{figure}
  \centering
  \includegraphics[height = 7cm]{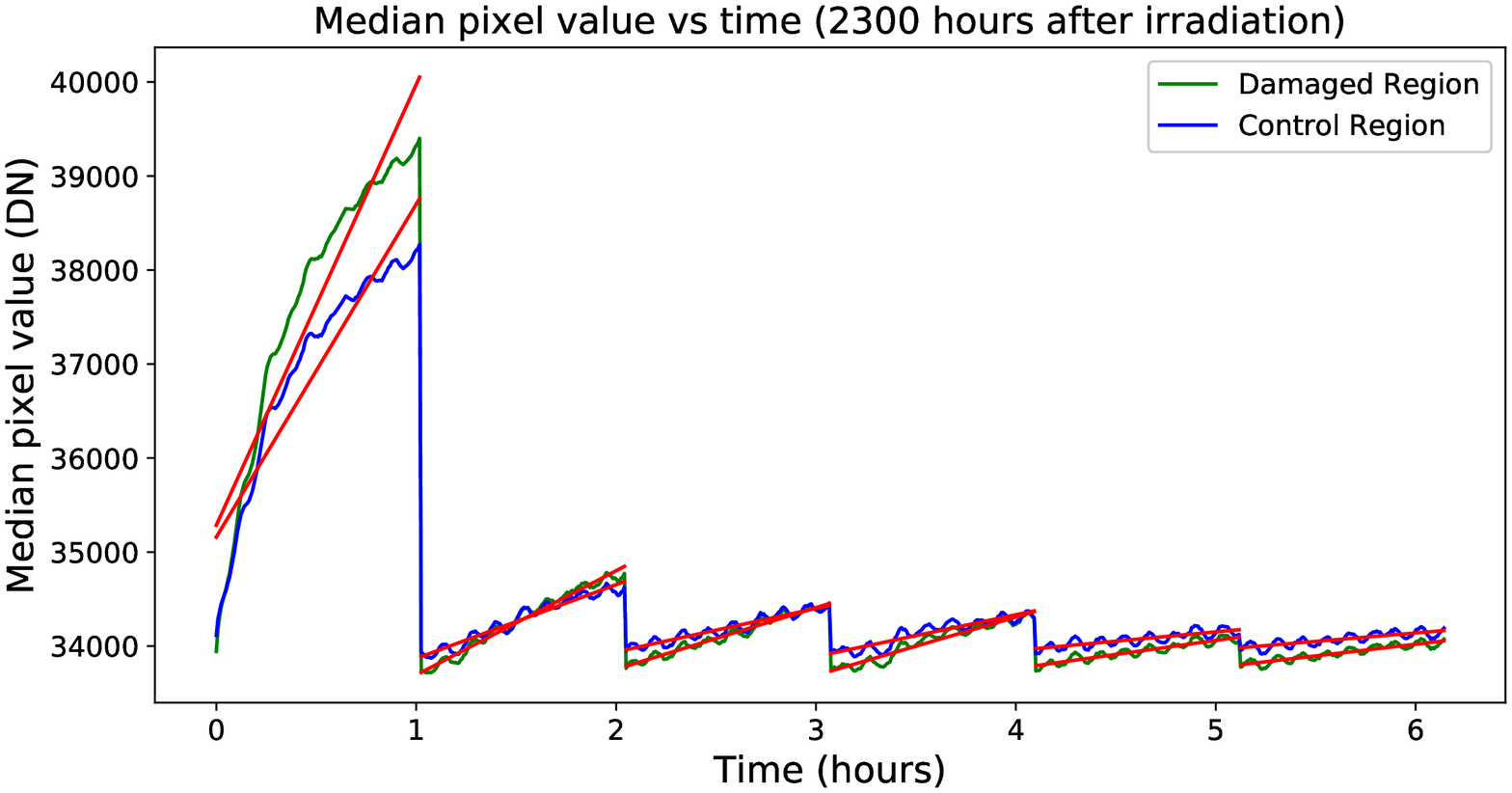}
  \caption{The dark current is compared between the damaged and control regions of the detector by performing a linear fit to the median pixel values of each image in a ramp. As expected, the dark current is higher in the damaged region. The small scale fluctuations are caused by the temperature control system oscillating around the set value of 150K. The changing slope from initial to final ramp indicates a long timescale charge-clearing effect as well. The dark currents cited in this paper are calculated once the slope has reached an equilibrium point after several hours.}
\label{fig:dark_current_ramps}
\end{figure}

\begin{figure}
\begin{subfigure}{.5\textwidth}
  \centering
  \includegraphics[height = 6cm]{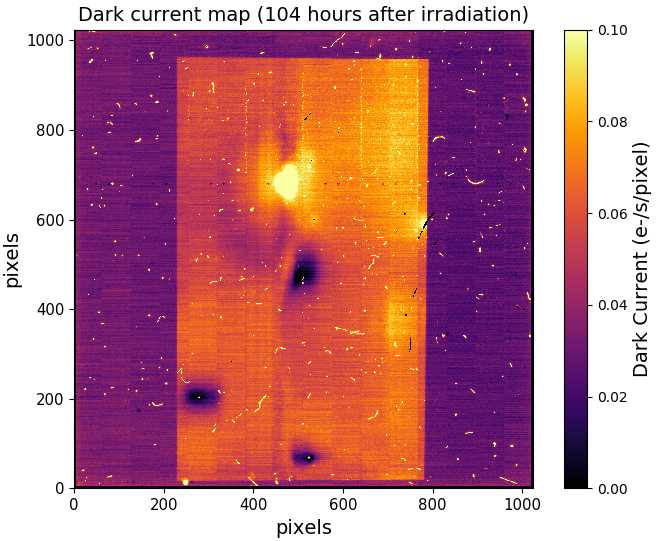}
\end{subfigure}
\begin{subfigure}{.49\textwidth}
	\centering
	\includegraphics[height = 6cm]{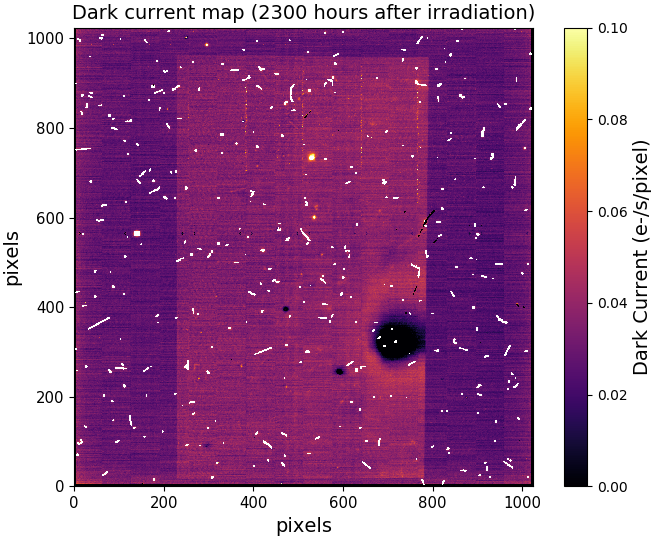}
\end{subfigure}
  \caption{A map of the dark current across the detector during a one hour ramp, taken 104 hours (Left) and 2300 hours (Right) after irradiation. The damaged region can be seen in the middle, with the control regions on either side. Cosmic rays appear scattered over the image, and the activation events described in Section \ref{sec:Activation Events} appear as bright and dark spots.}
\label{fig:dark_current_maps}
\end{figure}

\begin{figure}
  \centering
  \includegraphics[height = 6.3cm]{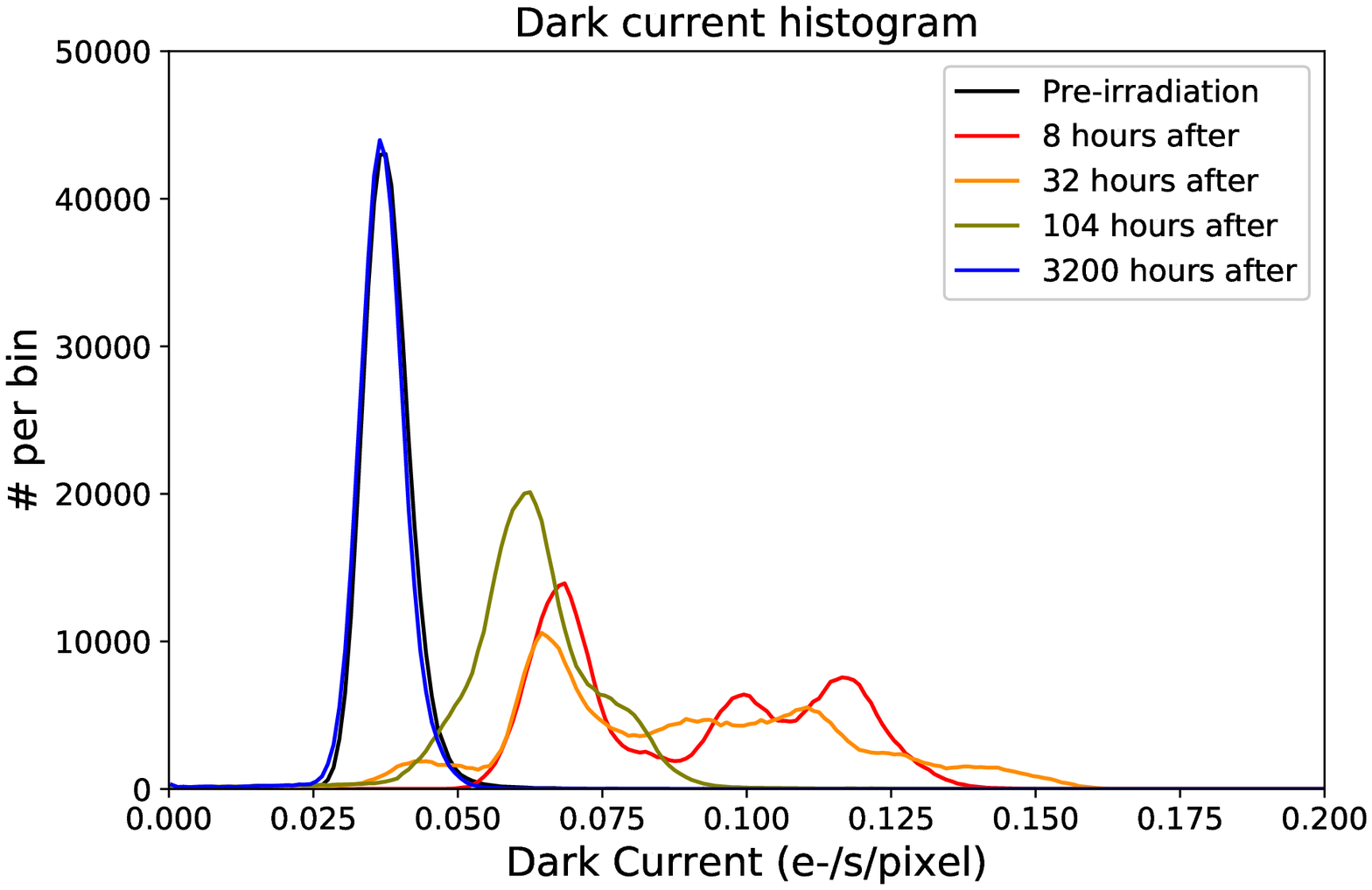}
  \caption{A dark current histogram of the damaged region is compared at multiple epochs after irradiation. As the detector anneals, the dark current appears to return to pre-irradiation levels and the distribution recovers its Gaussian profile. The strong tails are produced by the activation events discussed in Section \ref{sec:Activation Events}.}
\label{fig:dark_current_histogram}
\end{figure}

We measured the dark current by performing non-destructive reads of the detector during a series of one-hour ramps. No X-ray source was on while the images were being collected, and ramps containing many images were taken so that the accumulated dark signal could be distinguished from the noise floor, as shown in Figure \ref{fig:dark_current_ramps}. We assume that dark current and read noise are constant during each ramp, and perform a linear fit to the median pixel values of each image to determine the dark current. Pixels that show signs of having interacted with a cosmic ray by exhibiting a sudden, large increase in signal are automatically flagged and removed from the analysis, as are the eight adjacent pixels surrounding them. Dark current is also analyzed for each individual pixel, demonstrated in Figure \ref{fig:dark_current_maps}, and the results are histogrammed, as shown in Figure \ref{fig:dark_current_histogram}. After a large initial shift, the measured values of dark current appear to be returning to pre-irradiation levels as the detector anneals.

\subsection{Energy Resolution} \label{sec:EnergyResolution}
\begin{figure}
  \centering
  \includegraphics[height = 6cm]{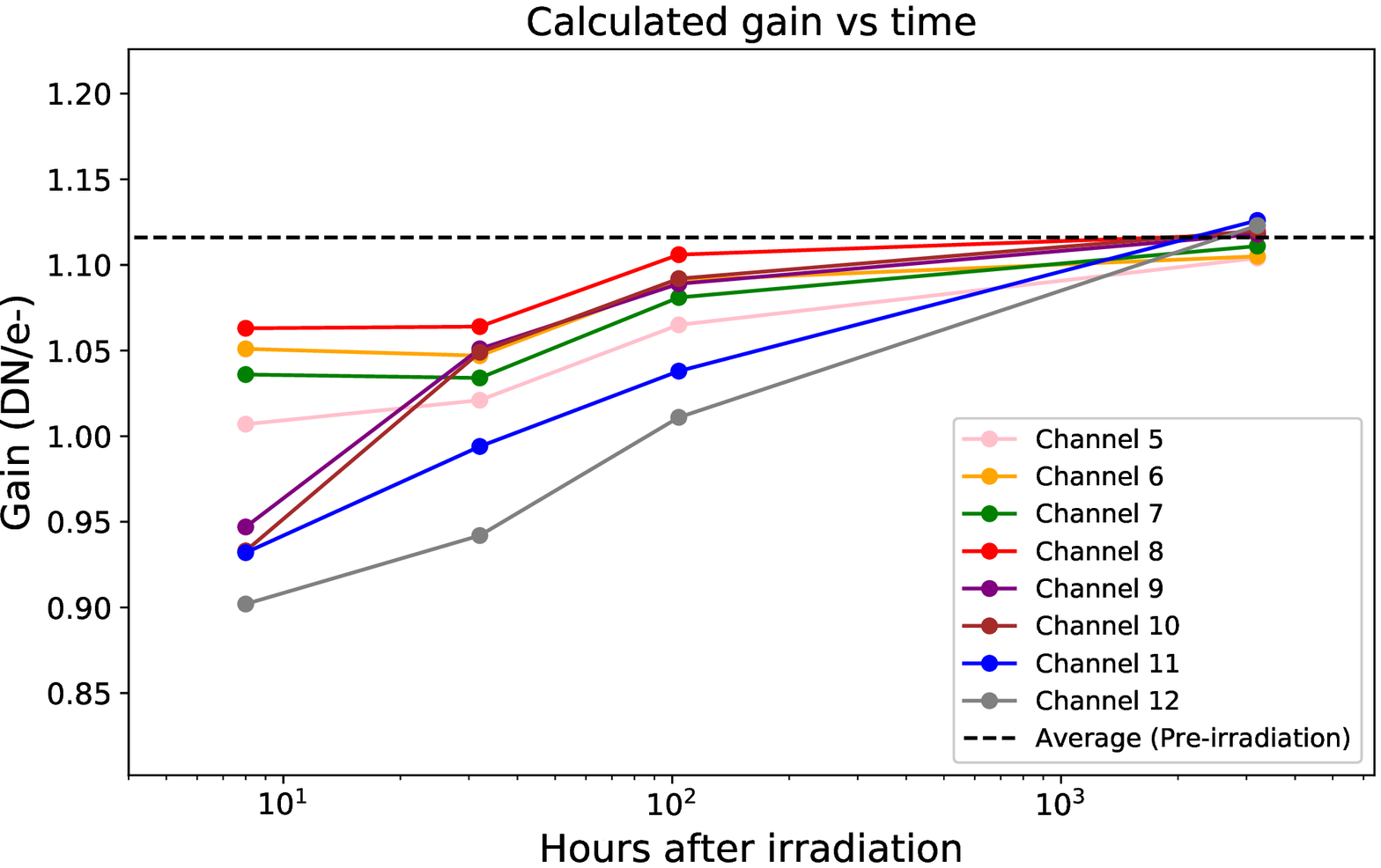}
  \caption{A plot of the calculated gain vs time at several multiple epochs after irradiation. After a large initial shift, gain appears to have returned to a nominal level.}
\label{fig:gain_vs_time}
\end{figure}

We measured the detector gain and energy resolution by utilizing a radioactive Iron-55 source mounted on a shutter system inside the vacuum chamber. After irradiation, a distinct shift in the DN value of the centroid peak was observed over the measurement period. It is uncertain whether this change represents an actual shift in the gain of the readout amplifiers in the ROIC, or a decrease in the number of electrons that are produced by each X-ray that are reaching the collection node. Interpreting this shift as an effective gain change, we plot the gain of each channel as a function of time in Figure \ref{fig:gain_vs_time}. Following annealing during several months of storage, the energy resolution was not significantly degraded from its pre-irradiation value of 6.9$\%$ at 5.9keV.

\subsection{Likely Activation Events} \label{sec:Activation Events}
In the days immediately following irradiation, we notice anomalous signals in our images with an occurrence rate of roughly 1-2 per hour. Similarly to cosmic rays, these events quickly impart a large amount of signal to a macroscopic area of the detector, often affecting hundreds to thousands of pixels at a time, but always within the damaged region of the detector. Affected pixels show a sudden jump in overall signal level and dark current, with severity depending on their proximity to the center of the event. After the charge is cleared by a detector reset at the end of the ramp, the affected pixels then show a negative trend in pixel value for the majority of the next ramp. This results in a measurement of apparent negative dark current during the ramp following the one when the event happened. Examples of these events appear as bright and dark spots in Figure \ref{fig:dark_current_maps}. After several months, the frequency of these events has decreased to $\sim$1 every 8 hours, which could be an indicator that they are radioactive in origin. This is likely a byproduct of residual activation of material in the detector, which is only still present due to the very large radiation dose being delivered over such a short time period. Because of the large number of affected pixels, histograms of the dark current contain significant non-Gaussian features, as demonstrated by Figure \ref{fig:dark_current_histogram}. 


\section{Conclusion and Future Work}
Initial analysis of the data taken post-irradiation indicate results that are consistent with experiments on similar devices; an initial degradation of detector characteristics, followed by a slow annealing process. After several months of storage, no measurable amount of performance degradation was observed in read noise, dark current, gain, or spectral resolution, as summarized in Table \ref{table:summary}. In the coming months, we will continue to evaluate detector performance before returning to Ohio University to carry out further irradiations at higher doses. By observing how these features change with time at various absorbed doses, we can gain a better understanding of the practical effects of annealing in these detectors. After characterizing the final equilibrium state of common detector features, we can determine the extent to which the current generation of HCDs can operate in a space-like environment without suffering radiation damage that would compromise mission integrity.

\section*{Acknowledgments}
We would like to thank the Physics Department at Ohio University and the staff of the Edwards Accelerator Lab for the generous use of their facilities and equipment, as well as providing the necessary guidance and expertise to ensure a successful experiment. This work was supported by NASA grants NNX16AO90H, NNX14AH68G, and 80NSSC18K0147. The work at the Edwards Accelerator Laboratory was supported in part by the Ohio University College of Arts and Sciences and the U.S. Department of Energy, Grant No. DE-FG02-88ER40387.


\newpage
\bibliography{Radiation_Experiment_Manuscript}{} 
\bibliographystyle{unsrt}
\end{document}